\documentclass[amsmath,amssymb,aps,prb,reprint,superscriptaddress,nobibnotes]{revtex4-1}

\usepackage{amsmath}
\usepackage{amsfonts}
\usepackage{amssymb}
\usepackage{graphicx}
\usepackage{upgreek}
\usepackage{color}
\usepackage{braket}
\usepackage{verbatim}
\usepackage{xr}

\makeatletter 
\renewcommand\@biblabel[1]{#1.} 
\makeatother

\definecolor{darkred}{rgb}{0.5, 0, 0}
\definecolor{darkgreen}{rgb}{0, 0.5, 0}
\definecolor{darkblue}{rgb}{0.1, 0.1, 0.7}

\usepackage[bookmarks=true,
            bookmarksnumbered=false,
            bookmarksopen=true,
            breaklinks=true,
            pdfborder={0 0 0},
            final=true,
            backref=none,
            colorlinks=true,
            linkcolor=darkblue,
            menucolor=darkblue,
            citecolor=darkblue,
            urlcolor=darkblue]{hyperref}

\newcommand{\micro}{${\upmu}$}

\newcommand{\upsub}[1]{_{\mathrm{#1}}}
\newcommand{\um}{$\,$\micro m}
\newcommand{\uev}{$\,$\micro eV}

\newcommand{\ZB}{\textit{Zitterbewegung}}

\begin{document}


\title{Observation of \ZB{} in photonic microcavities}

\author{Seth Lovett}
\author{Paul M. Walker}
\email{p.m.walker@sheffield.ac.uk}
\affiliation{Department of Physics and Astronomy, University of Sheffield, S3 7RH, Sheffield, UK}

\author{Alexey Osipov}
\author{Alexey Yulin}
\affiliation{Department of Physics and Technology, ITMO University, St. Petersburg, 197101, Russia}

\author{Pooja Uday Naik}
\author{Charles E. Whittaker}
\affiliation{Department of Physics and Astronomy, University of Sheffield, S3 7RH, Sheffield, UK}

\author{Ivan A. Shelykh}
\affiliation{Science Institute, University of Iceland, Dunhagi 3, IS-107, Reykjavik, Iceland}
\affiliation{Department of Physics and Technology, ITMO University, St. Petersburg, 197101, Russia}

\author{Maurice S. Skolnick}
\author{Dmitry N. Krizhanovskii}
\affiliation{Department of Physics and Astronomy, University of Sheffield, S3 7RH, Sheffield, UK}


\begin{abstract}
We present and experimentally study the effects of the photonic spin-orbit coupling on real space propagetion of polariton wavepackets in planar semiconductor microcavities and polaritonic analogs of graphene. In particular, we demonstrate the appearance of an analog \ZB{} effect, a term which translates as 'trembling motion' in english, which was originally proposed for relativistic Dirac electrons and consists of the oscillations of the center of mass of a wavepacket in the direction perpendicular to its propagation. For a planar microcavity we observe regular \ZB{} oscillations whose amplitude and period depend on the wavevector of the polaritons. We then extend these results to a honeycomb lattice of coupled microcavity resonators. Compared to the planar cavity such lattices are inherently more tuneable and versatile, allowing simulation of the Hamilitonians of a wide range of important physical systems. We observe an oscillation pattern related to the presence of the spin-split Dirac cones in the dispersion. In both cases the experimentally observed oscillations are in good agreement with theoretical modelling and independently measured bandstructure parameters, providing strong evidence for the observation of \ZB{}.
\end{abstract}

\maketitle


\section*{Introduction}

The study of analogs to effects appearing in the domain of high energy physics is among the trends of the modern condensed matter physics. In this connection, Fabry-Perot optical microcavities, structures where the internal spinor wavefunction can be directly imaged via photon tunnelling through the mirrors, are of particular interest.
Such cavities generally support ballistic propagation of two polarisation states of light and have an inherent polarisation-wavevector coupling, the TE-TM splitting, which is an equivalent of spin-orbit coupling (SOC) \cite{Shelykh2010}. They may also support birefringent polarisation splitting, all of which features have allowed observation of a range of important physical effects such as optical spin-Hall effect~\cite{Leyder2007}, emergence of monopoles \cite{Hivet2012} and onset of the non-Abelian gauge fields \cite{Polimeno2021,Bieganska2021}.
In semiconductor microcavities, quantum wells (QWs) can be added resulting in formation of composite part-light part-matter quasiparticles called exciton-polaritons. Their unique properties, related to the extremely small effective mass (about five orders of magnitude less then the mass of a free electron), high sensitivity to magnetic fields, giant nonlinear optical interactions, and possibility for optical amplification, have allowed demonstration of optical condensates with macroscopically large coherence length (in the mm scale) \cite{Ballarini2017}, formation of acoustic black holes and Hawking effect \cite{Jacquet2020}, observation of anomalous Hall drift~\cite{Gianfrate2020}, and many other effects.
In this work we use such semiconductor microcavities but focus on the properties coming from the photonic constituent of polaritons, and so we demonstrate the relevant fundamental principles using highly photonic ($>98$\%) polaritons.

Furthermore, there has been much recent work on engineering the bandstructure of microcavities by imposing a laterally varying optical potential~\cite{Milicevic2017,Whittaker_2018,Alyatkin2020}. This has allowed study of flat bands in Lieb lattice potentials~\cite{Whittaker_2018}, topological physics~\cite{Gao2018,Klembt2018,Solnyshkov2021} and engineering of Dresselhaus SOC for photons~\cite{Zhang2020,Whittaker_2020}. Such lattices, as well as being substantially more tuneable via the added in-plane degrees of freedom, can be used to build photonic analogs of a wide variety of physically important Hamiltonians based on Bose-Hubbard type models~\cite{Schneider_2016}. Combined with the other favourable properties of polaritons discussed above this creates a very wide perspectives for polaritonic simulation.

One of the textbook examples of quantum relativistic effects is \ZB{}. It consists of an oscillatory motion of a propagating wavepacket transverse to its ballistic trajectory despite the absence of transverse forces~\cite{Zawadzki_2011}. It was first predicted by Schr{\"o}dinger for the motion of free electrons governed by the Dirac equation~\cite{Barut_1981} and appears due to interference between positive and negative energy states of a spinor (two component) system, enabled by the coupling of internal (spin) and external (momentum) degrees of freedom.

In addition to free relativistic electrons the effect is predicted for electrons in crystals with Rashba and Dresselhaus SOC~\cite{Schliemann_2005,Schliemann_2006,Winkler_2007}. The predicted high frequency and low amplitude of the oscillations for the vacuum case, and the difficulty of observing single electrons in solids, make experimental observation highly challenging~\cite{Zawadzki_2011}. This has led to a search for analagous systems in which the observation can be made. Gerrtisma et. al.~\cite{Gerritsma_2010} performed a quantum simulation of the Dirac equation using a single trapped ion. Measurement of the transverse position was made indirectly since most observables cannot be directly measured in ion trap experiments. High frequency oscillating currents were also observed in the motion of spin-polarised electrons in a doped semiconductor device~\cite{arxiv.1612.06190}. In optics, the \ZB{} was observed in arrays of coupled waveguides where the internal degree of freedom mimicking spin was introduced by having two slightly different waveguides per unit cell~\cite{Dreisow_2010}. In that case the energy separation of positive and negative components was fixed by the geometry of the waveguides rather than being continuously tuneable. Thus different lattices had to be used to examine different parameters. Furthermore, the \ZB{} had to be detected indirectly using the fluorescence induced by the intensity of the light inside the waveguides. Microcavities, by contrast, allow direct imaging and excitation of the internal wavefunction and addressing of different energy separations using pump laser incidence angle and frequency.

Whereas \ZB{} is usually understood as an oscillation occurring without external forces it is worth noting that zig-zag oscillations have also been observed in microcavities with a transverse trapping potential~\cite{Rozas2020}. A possible explanation in terms of \ZB{} was suggested~\cite{Rozas2021} but still awaits confirmation through detailed analysis and comparison with theory to rule out the effects of the transverse potential.
Similarly, it was recently shown that polariton SOC contributes to the periodicity of transverse oscillations of a condensate in an etched ring trap, alongside the contribution from condensate motion in the transverse trapping potential~\cite{Yao2022}. Since the SOC underpins \ZB{} in planar structures its contribution may be interpreted as a manifestation of \ZB{} in the ring traps. However, no direct observation of \ZB{} was possible. In general, oscillations in structures with transverse trapping potentials are hard to attribute to \ZB{} since there are alternative explanations such as interference between multiple transverse modes~\cite{Sich2018}.

So far \ZB{} has not been directly observed in the microcavity structures where polaritons can be formed and which allow the wide range of optical analogs discussed above, although it has recently been theoretically predicted in both planar microcavities~\cite{Sedov_2018} and honeycomb microcavity lattices at wavevectors close to the Dirac point~\cite{Whittaker_2020}. 
In this work we bridge this gap between theory and experiment and report observation of \ZB{} in both types of structures. For the planar cavity we demonstrate tuning of the \ZB{} period by varying the incidence angle and propagation direction in the birefringent cavity. In the case of wavevectors near the Dirac point in a honeycomb lattice the ability to engineer the bandstructure allows observation of smaller period \ZB{} while retaining an observable amplitude.
The \ZB{} essentially arises from interferences between the two components of the spinor wavefunction. The propagating beam has finite spatial width and hence finite width in momentum (angular) space. Since the polarisation states depend on the momentum through the spin orbit coupling, different angular components of the beam have different relative amplitudes of the spinor components, and these can evolve with propagation.
As discussed above we experimentally demonstrate the essential principles of the effect in the highly photonic regime, but polaritons can be made more excitonic by simple tuning of the photon-exciton detuning, which opens up wider perspectives.

\section*{Results}

\begin{figure}
\includegraphics[width=\columnwidth]{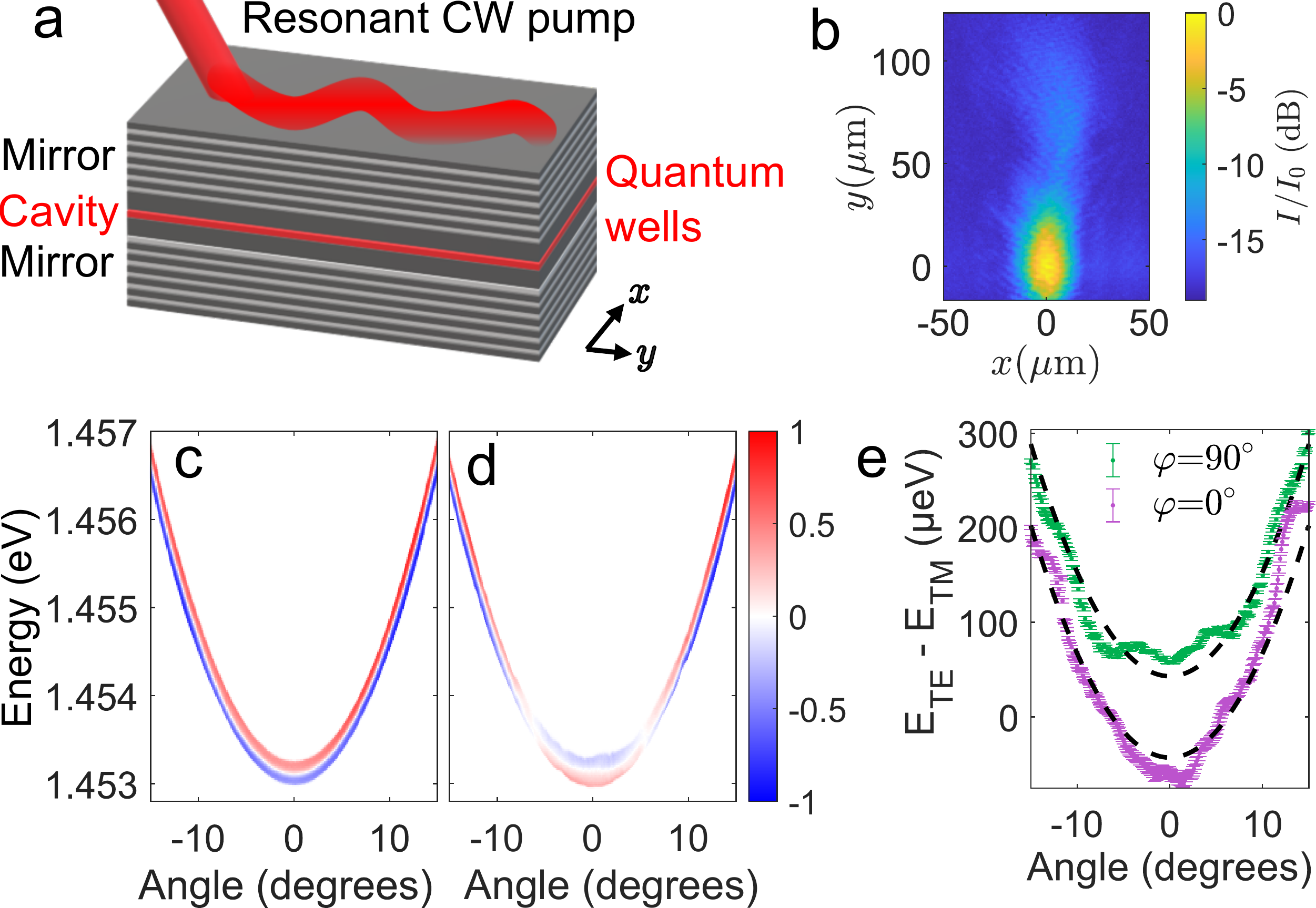}
\caption{(a) Schematic of the planar cavity structure resonantly excited at one point with the photons propagating away along the cavity. (b) Intensity of the photon field in the cavity when excited resonantly vs. $x$ and $y$. Colour scale gives the intensity $I$ relative to the peak intensity $I_0$ in decibel units. The zero of the $y$-axis is defined as the point of peak intensity vs. $y$. (c) Angle and polarisation resolved photoluminescence spectrum showing the dispersion relation $E\left(k_{y}\right)$ at a fixed $k_x=0$ for the case where the birefringent crystal principle axis $y^{\prime}$ ($\varphi=90^{\circ}$, see supplementary section 1) is parallel to the direction $y$ along which the polaritons are injected in the resonant excitation experiment. (d) As panel (c) but for the case where $x'$ is parallel to $y$ ($\varphi=0^{\circ}$). In (c) and (d) the colour scale indicates the polarisation degree $\left(I_{x} - I_{y}\right)/\left(I_{x} + I_{y}\right)$ with red indicating $x$ polarisation and blue indicating $y$ polarisation ($I_{x}$ and $I_{y}$ are the intensities in the $x$ and $y$ polarisations). Points with total intensity $\left(I_{x} + I_{y}\right)$ less than 0.20 of the peak have been set to white since the polarisation degree is not well defined for low intensities. (e) The energy splitting between the TE and TM polarisations for the two values of $\varphi$. Points show the values extracted by fitting Lotentzian peaks to the data in panels (c) and (d). Dashed black curves are the fits described in the main text.}
\label{fig:schematic}
\end{figure}

We begin with the case of the planar microcavity. In these structures a two-wavelength thick cavity layer is enclosed between two Bragg mirrors (periodically repeating stacks of quarter-wave layers of two different materials), as illustrated in the schematic in Fig~\ref{fig:schematic}(a). In our case the cavity is made of GaAs, the mirror materials are GaAs and Al$_{0.85}$Ga$_{0.15}$As, and the structure was grown by molecular beam epitaxy. Three In$_{0.04}$Ga$_{0.96}$As QWs are embedded in the cavity. The energy detuning between the QW excitons and the the cavity photons is more than 20 meV.

The polarisation and angle dependent reflection of the cavity mirrors leads to a wavenumber dependent energy splitting (TE-TM splitting) of the cavity modes having electric and magnetic fields transverse to the wavevector. This combines with a slight optical birefringence \cite{Martin2004,Amo2005,Tercas2014} resulting in a complicated splitting between linear polarization states. For small in-plane wavevector components the Hamiltonian of the system in the basis of circular polarized states reads \cite{Tercas2014}:
\begin{eqnarray}
\hat{H}=\left(\begin{array}{cc}
 \frac{\hbar^2 k^2}{2m}    &  \frac{\Omega}{2} -\beta(k'_x-ik'_y)^2\\
  \frac{\Omega}{2} -\beta(k'_x+ik'_y)^2   & \frac{\hbar^2 k^2}{2m}
\end{array}\right),
\end{eqnarray}
where $m$ is the effective mass of the polaritons, $k'_x$ and $k'_y$ are the in-plane wavevector components of the photons in the sample reference frame where $x'$ is the fast axis (see Supplementary Fig. 1), $k^2 = {k^{\prime}_x}^{2} + {k^{\prime}_y}^{2}$, and parameters $\Omega$ and $\beta$ describe the values of the k-independent optical birefrigence 
and TE-TM spliting respectively. The parameter $\beta$ is related to the difference of the longitudinal and transverse masses of the photons $m_l$ and $m_t$ as \cite{Flayac_2010} 
\begin{equation}
\beta=\frac{\hbar^2}{4}\left(\frac{1}{m_t}-\frac{1}{m_l}\right)
\end{equation}
To clarify the notation, $m_t$ and $m_l$ are the masses of the TE and TM polarised photons. Note that we define $\beta$ with opposite sign compared to the definitions in Ref. \cite{Tercas2014} and Ref. \cite{Flayac_2010} but that this does not affect the physics. The corresponding dispersions of the two photon branches split in linear polarisations read:
\begin{equation}
    E_\pm=\frac{\hbar^2 k^2}{2m}\pm\sqrt{\beta^2k^4-\beta\Omega k^2\cos2\varphi+\frac{\Omega^2}{4}},
\end{equation}
where $\varphi$ is the in-plane angle between the wavevector $\mathbf{k'}$ and the $x'$ axis of the crystal. With these definitions, the energy of the TE polarised mode (electric field perpendicular to $\mathbf{k'}$) increases faster with $k$ than the TM mode for positive $\beta$, and at $k=0$ the mode polarised along $x'$ has higher energy for positive $\Omega$. Note, that the combination of birefrigence and TE-TM splitting leads to a clear in-plane anisotropy of the dispersions, which cross for $\varphi=0$ at $k=\sqrt{\Omega/\left(2\beta\right)}$. This will now be seen experimentally as we present the basic characterisation of the sample.

The experiments in this paper were performed at approximately 10 K temperature in a continuous-flow cold-finger cryostat. The energy vs. wavevector dispersion relations $E(k_y)$ at $k_x=0$ were measured by angle and polarisation resolved photoluminescence (PL) spectroscopy and can be seen in Fig.~\ref{fig:schematic}(c) and (d). Note that $x$ and $y$ are coordinates in the laboratory reference frame (see Fig.~\ref{fig:schematic}(a)). In both figures the angle $\theta$ on the horizontal axis gives the wavevector $k_y=k_0\sin{\theta},k_x=0$, where $k_0=2\pi/\lambda$. Fig.~\ref{fig:schematic}(c) shows the case where the sample is rotated such that $\mathbf{k}=k_{y}\mathbf{\hat{y}}$ is parallel to $y'$ ($\varphi=90^{\circ}$). Two branches with different polarisation are visible. As expected there is a splitting at $k=0$ resulting from the birefringence and the splitting increases with $k$ due to the the TE-TM splitting. Fig.~\ref{fig:schematic}(d) shows the case where the sample is rotated such that $\mathbf{k}=k_{y}\mathbf{\hat{y}}$ is parallel to $x'$ ($\varphi=0^{\circ}$). In this case the dispersions cross at 5.7$^{\circ}$ ($k=$ 0.73 $\upmu$m$^{-1}$), again as expected. Note that at the crossing point the polarisation degree is low because the contributions from the two branches have similar intensity but are not coherent with one another owing to the spontaneous nature of photoluminescence emission. For each measured dispersion we extract the energy at the peak intensity for each angle and for each polarisation and fit the resulting energy vs. angle curves to find the parameters describing the sample. We find $\Omega=$ 43 $\pm$ 19 \uev{}, $\beta=$ 33.6 $\pm$ 3.5 \uev{} \um$^{2}$. The energy at $k=0$ (averaged between the two dispersion) is 1.4531 eV and $\hbar^2/\left(2m\right)=$ 947.5 \uev{} $\upmu$m$^{2}$.

We now proceed to measure the \ZB{} effect in the sample. We excited the cavity resonantly using a tuneable continuous wave Ti:Sapphire laser as illustrated in Fig.~\ref{fig:schematic}(a). The laser energy and incident angle were set to match points on the dispersion allowing efficient injection of photons into the cavity. The laser spot incident on the sample was circular with full-width-at-half-maximum (FWHM) 15\um{}. The size of the Fourier transform of this spot, e.g. its size vs. wavenumber $k$ was sufficient to efficiently excite both polarisation branches. The excitation polarisation was circular to ensure equal excitation of both branches. Light was collected from the opposite side of the sample to the excitation (transmission geometry) and the total intensity recorded by a thermo-electrically cooled CCD camera. An example of the recorded intensity pattern is shown in Fig.~\ref{fig:schematic}(b). The highest intensity part peaking at $y=0$ \um{} is related to the incident laser spot. Since we excite at finite angle where the group velocity (slope of the dispersion) is finite the photons in the cavity propagate away from the excitation spot in the positive $y$ direction, decaying by emitting photons through the Bragg mirrors towards the detector. For any value of $y$ we can take a slice along $x$ and find the value of $x$ at which the intensity is maximum. To minimise the effects of scatter of the data points it is better to define the center of intensity according to

\begin{equation}\label{eq:com}
x\upsub{c}(y) = \frac{\iint_{-\infty}^{\infty} x\cdot I(x,y) \cdot dx}{\iint_{-\infty}^{\infty} I(x,y) \cdot dx}
\end{equation}

\begin{figure}
\includegraphics[width=\columnwidth]{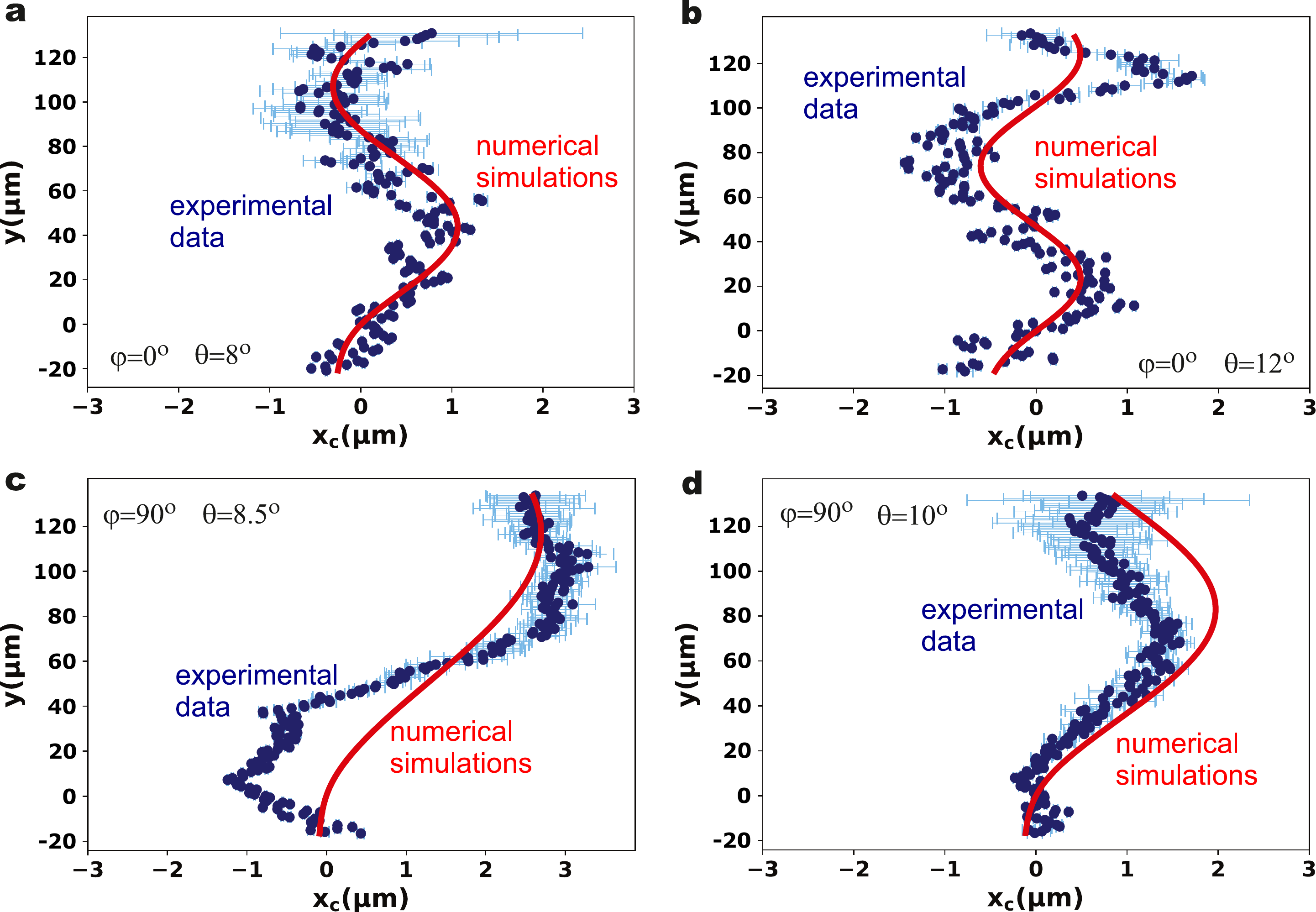}
\caption{Center of intensity $x\upsub{c}$ vs $y$ for resonant excitation at two different angles and for two orientations of the sample with respect to the propagation direction $y$. (a) and (b) are for the case where the principle axis $y'$ is aligned along $y$ and correspond to the dispersion curve in Fig.~\ref{fig:schematic}(c). (c) and (d) are for the case where the principle axis $x'$ is aligned along $y$ and correspond to the dispersion curve in Fig.~\ref{fig:schematic}(d). Blue points correspond to the experimental data, red solid line gives the theoretical fit. The zero of the $y$-axis is defined as the point of peak intensity vs. $y$. The zero of the $x$ axis is defined such that the theoretical $x\upsub{c}\left(y=0\right)=0$. Error bars are plotted between $\pm\sigma$ where $\sigma$ is the standard deviation calculated as discussed in Supplementary section 4. Error bars smaller than the data point symbols are not plotted.}
\label{fig:fig2}
\end{figure}

In Fig.~\ref{fig:fig2} we plot $x\upsub{c}(y)$ for different incidence angles and orientations of the sample. Figs.~\ref{fig:fig2}(a) and (b) show the cases for 8$^{\circ}$ and 12$^{\circ}$ angles of incidence respectively and where we inject light along the $y'$ axis of the cavity ($\varphi=0$). The blue points give the experimentally extracted trajectory of the wavepacket. Clear oscillations of $x\upsub{c}$ are visible with increasing distance $y$ away from the excitation spot. We discuss sources of uncertainty in $x\upsub{c}$ in Supplementary section 4.
For the \ZB{} effect the period of the oscillation $L$ corresponds to the frequency separation of the positive and negative branches and is expected to vary with excitation angle~\cite{Sedov_2018}.
Qualitatively, at larger angle we expect higher frequency (shorter period) oscillations as the TE-TM splitting between the branches increases (see Fig.~\ref{fig:schematic}e).
As expected we see in the experiment that the period becomes shorter at higher angles.
We simulated the expected trajectory of the wavepackets using a model of the evolution of the spinor wavefunction accounting for both TE-TM splitting and birefringence (see Supplementary section 1). The red curves in Figs.~\ref{fig:fig2}(a) and (b) show the theoretical evolution of $x\upsub{c}$ vs. $y$. The parameters $\Omega =$ 28.8 $\upmu$eV$\cdot\upmu$m$^2$ and $\beta =$ 32.45 $\upmu$eV$\cdot\upmu$m$^2$ for the simulation were obtained by fitting the experimental oscillations. Both the period and amplitude of the simulated trajectories are in excellent agreement with the experimental points.
Importantly, the values of $\beta$ and $\Omega$ found from fitting the oscillations are in good agreement with the values found independently from fitting the dispersion relations ($\beta$ agrees within 0.33 of the uncertainty, $\Omega$ agrees within 0.75 of the uncertainty), as discussed above.
We then rotated the sample and measured the oscillations for the case where we inject light along the $x'$ axis of the cavity ($\varphi=90^{\circ}$). As discussed above, this changes the nature of the dispersion, in particular the splitting between the two polarisation branches for a given angle. The data is shown in Figs.~\ref{fig:fig2} (c) and (d) for incidence angles 8.5$^{\circ}$ and 10$^{\circ}$ respectively. We used exactly the same parameters as in the $\varphi=0^{\circ}$ case to simulate the trajectory of the light and obtain semi-quantitative agreement between the experimental points and the theory, without further fitting.
The agreement of the amplitude is slightly less good than in the $\varphi=0^{\circ}$ case due to proximity to the crossing of the two polarisation branches (Fig.~\ref{fig:schematic}(d)) at 5.7$^{\circ}$. At this point there is no splitting and the period becomes singular, with the oscillations being correspondingly more sensitive to the exact parameters at angles close to the singular point. 

Finally, the \ZB{} effect arises due to interference between the two polarisation components, e.g. it is fundamentally a spinor effect. From theory we expect that the oscillations in the total intensity pattern should disappear if only one polarisation branch is excited. We tested this in two ways, described in more detail in Supplementary section 2.
First, we tuned the excitation angle so that one branch was preferentially excited and saw that the amplitude of the oscillations gradually reduced and disappeared. Second, we excited the system with light linearly polarised parallel to one polarisation branch and perpendicular to the other and did not detect oscillations. 

\begin{figure}
\includegraphics[width=\columnwidth]{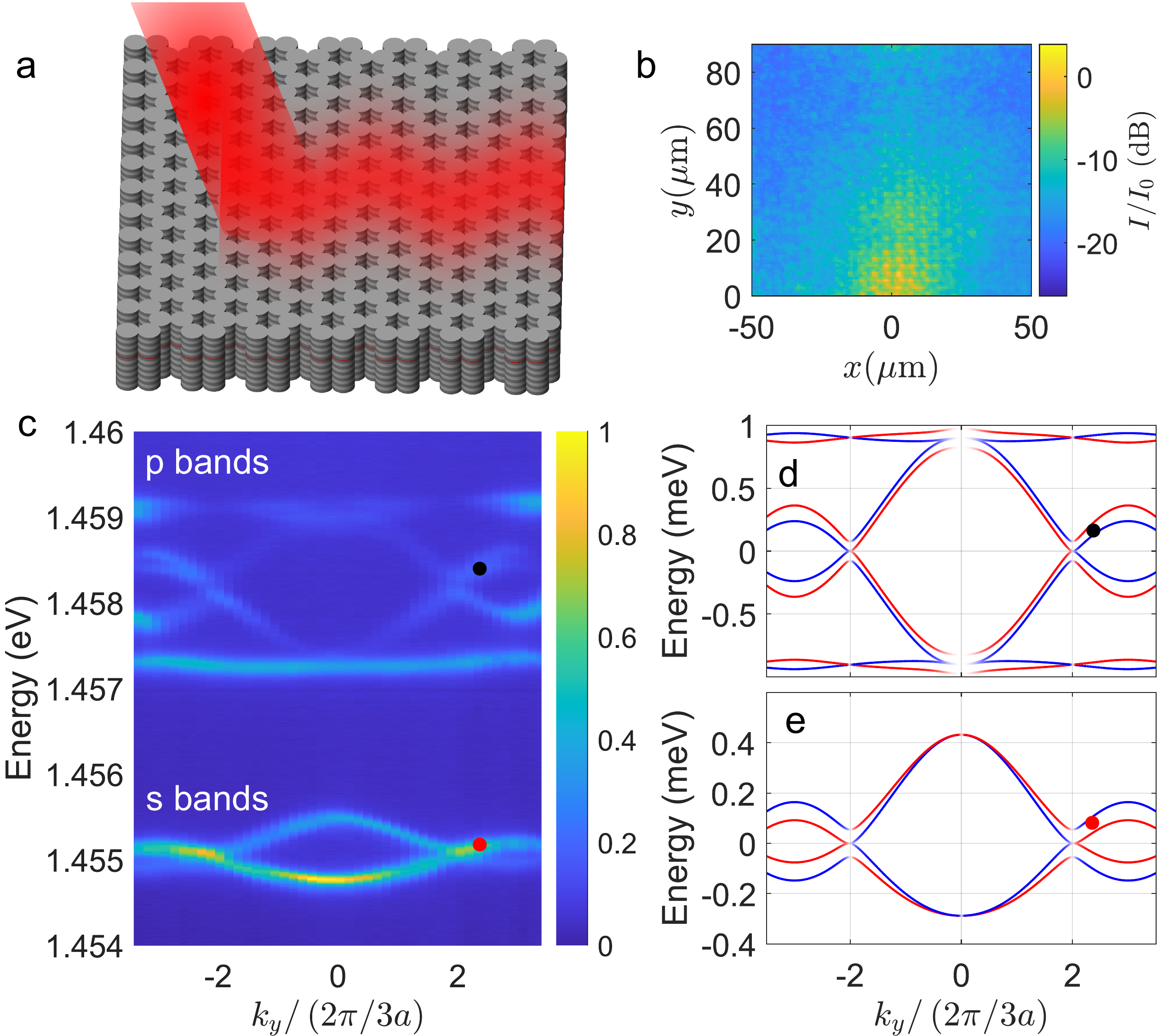}
\caption{(a) Schematic of the honeycomb lattice structure resonantly excited at one point with the photons propagating away along the structure. (b) Intensity vs. $x$ and $y$ recorded for excitation at the black point. Colour scale gives the intensity $I$ relative to the peak intensity $I_0$ in decibel units. The zero of the $y$-axis is defined as the edge of the lattice where the pump spot is incident. (c) Dispersion relation of the honeycomb lattice measured by angle-resolved photoluminesce spectroscopy without polarisation resolution. Colour scale gives the intensity relative to the peak value. The red and black points mark the excitation energies and corresponding incidence wavevectors at which the \ZB{} effect is studied. (d-e) Dispersion of the p-bands (d) and s-bands (e) calculated by a tight binding model with parameters from Ref.~\cite{Whittaker_2020}.
Red and blue curves represent states linearly polarised along $x$ and $y$ respectively.}
\label{fig:fig3}
\end{figure}

The planar microcavity, with its simple structure and tuneability through excitation angle, is a good platform for proof of principle demonstrations. Lattices formed by etching the microcavity to introduce a lateral pattern allow a wide variety of physically important bandstructures to be simulated and are highly tunable via the lateral patterning. They are thus an important complementary platform which open up a wide perspective for future studies. We therefore studied \ZB{} in a honeycomb lattice, a photonic analogue of graphene. The lattice is formed by etching the planar cavity into air post cavities. These so-called pillar microcavities support discrete energy states localised in all three spatial dimension. By overlapping neighbouring pillars (see Fig.~\ref{fig:fig3}) we allow photons to tunnel from one pillar to another, allowing the discrete states to hybridise and form energy bands. The structure we study here is a honeycomb lattice, which is a triangular lattice with 2 pillars per unit cell. The diameter of the pillars is 3\um{}, the center to center spacing of adjacent pillars is $d=2.8$\um{}, the lattice periodicity is $a=d\sqrt{3}$, the lattice vectors are $a\left(\sqrt{3}/2,\pm 1/2\right)$, and the reciprocal lattice vectors are $\left(2\pi/a\right)\left(1/\sqrt{3},\pm 1\right)$.

In Fig.~\ref{fig:fig3}(a) we show a schematic diagram of the honeycomb lattice structure being excited by the laser beam, resulting in propagation of the light in the cavity accompanied by transverse oscillations. An example of an experimentally measured intensity pattern is shown in Fig.~\ref{fig:fig3}(b) and is equivalent to the pattern seen in the planar case (Fig.~\ref{fig:schematic}(b)) except that it results from propagation of photons in the bands of the lattice rather than those of the planar cavity. The bandstructure of the lattice can be seen in the angle resolved photoluminescence spectrum of Fig.~\ref{fig:fig3}(c). Two distinct sets of bands separated by a band-gap can be seen. The lowest set of bands, labelled 's bands', are composed of the lowest energy states of the individual pillars, resembling lowest order Hermite-Gauss modes~\cite{Whittaker_2020}. The next highest set of bands, labelled 'p bands', are composed of the first order Hermite-Gaussian modes of the individual pillars. For the purpose of this work the essential difference between the two sets of bands is that the polarisation dependent tunnelling rates from pillar to pillar are different resulting in different splittings and different group velocities (Fermi velocity) close to the Dirac points (at $k_x=0$, $k_y=4\pi/\left(3a\right)$). We therefore expect different \ZB{} period and amplitude close to the Dirac points in either band. 

The red and black dots on Fig.~\ref{fig:fig3}(c) mark the energy and wavevector at which we resonantly excite the bands to measure the \ZB{}. As in the planar case we equally excite the two bands. More detailed plots of the bandstructure are given in Figs.~\ref{fig:fig3}(d) and (e) for the p and s bands respectively. These are obtained from a tight binding model using parameters from Ref.~\cite{Whittaker_2020} where the same lattice was studied extensively, a detailed analysis of polarisation splitting was performed, and the model was fit to the dispersion. The red and blue colour of the lines denote $x$ and $y$ polarised waves respectively and the dots (red for s-bands and black for p-bands) denote the energy and wavevector of excitation, slightly detuned from the Dirac points. Around these excitation points the energy splitting $\Delta E$ between the bands is 65\uev{} for the s-bands and 96\uev{} for the p-bands. The wavenumber splitting $\Delta k_y$ between the two bands at the fixed laser energy is $0.49\times 2\pi/\left(3a\right)$ for the s-bands and $0.25\times 2\pi/\left(3a\right)$ for the p-bands.

\begin{figure}
\includegraphics[width=\columnwidth]{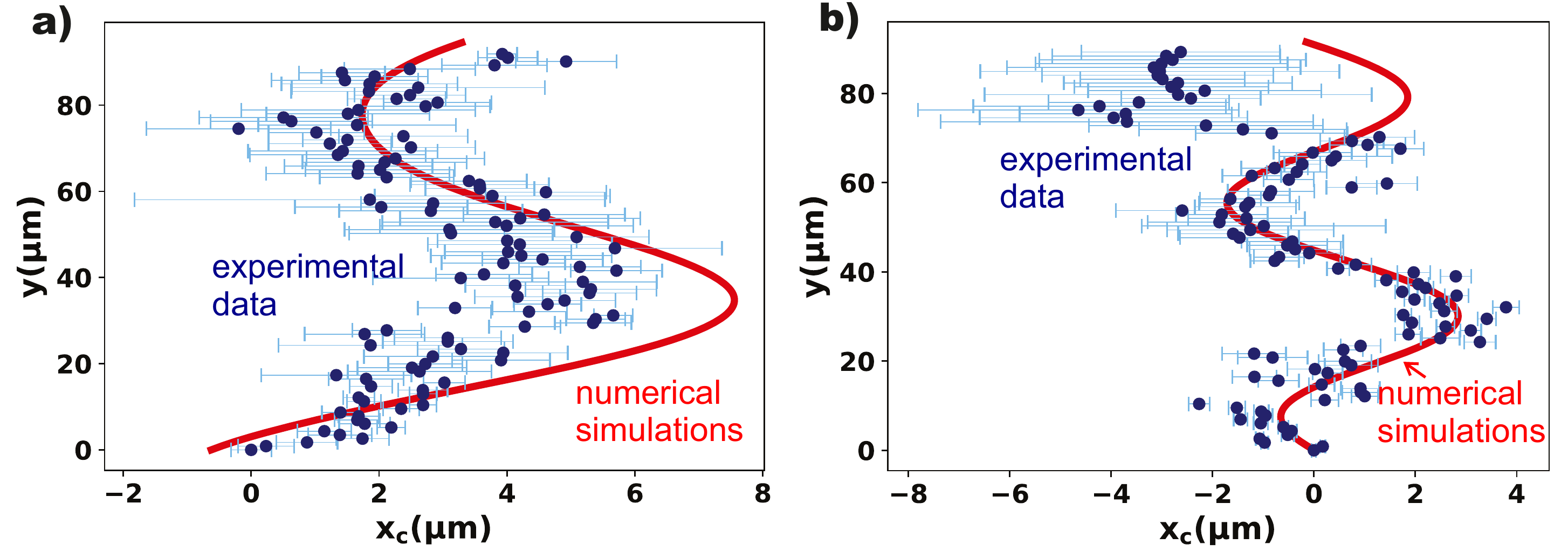}
\caption{Center of intensity $x\upsub{c}$ vs. propagation distance $y$ for resonant excitation at two different angles. Blue points give the experimental data. Red solid line gives the theoretical fit. (a) Case for excitation of the p-band states, corresponding to black circle in Fig.~\ref{fig:fig3}(d). (b) Case for excitation of the s-band states, corresponding to red circle in Fig.~\ref{fig:fig3}(e). The zero of the $y$-axis is defined as the edge of the lattice where the pump spot is incident.}
\label{fig:fig4}
\end{figure}

As in the planar case we use Eqn.~\ref{eq:com} to extract $x_c$ vs. $y$ from the intensity pattern of the propagating photons. Figs.~\ref{fig:fig4}(a) and (b) show the extracted trajectories of the photons for the p and s bands respectively. Oscillations of $x_c$ are visible and we see that the period of oscillation is shorter for the s-bands.
The period is expected to scale as $L=2\pi/\Delta k_y$. Therefore the shorter period for the s-band is consistent with the larger $\Delta k_y$ near the excitation point for the s-band.
We calculated the theoretical trajectory for the photons in a similar way as for the planar case, but using evolution equations for states near the Dirac point of the honeycomb lattice\cite{Nalitov_2015, Whittaker_2020} (see Supplementary section 3). Parameters for the modelling were taken from Ref.~\cite{Whittaker_2020} where the same lattice was studied extensively and the dispersion relation was fitted. Semi-quantitative agreement is obtained between the theory and experiment.

\section*{Discussion}

In the experiments on the planar cavity clear oscillations of the center of mass with increasing propagation distance were observed. The oscillation period decreased with increasing splitting between the branches and the oscillations disappeared when only one branch was excited, as expected. The expected trajectory of the center of mass was also calculated using a model of the evolution of the spinor wavefunction. The model reproduced the experimental period and amplitude using parameters in good agreement with those obtained independently from the dispersion curve measurements. The amplitude of \ZB{} oscillations is in general a complicated function of sample and excitation parameters but to a good approximation is governed by the experimentally chosen excitation angle~\cite{Sedov_2018} and, unlike the period, is independent of the TE-TM splitting and birefringence parameters $\beta$ and $\Omega$. The fact that both the observed period and amplitude agree well with the model provides strong evidence that the oscillations we observe experimentally are indeed the real transverse oscillations of the wavepacket in the cavity, which is the \ZB{}. We also studied \ZB{} in honeycomb lattices, where clear oscillations were once again observed with period and amplitude in good agreement with the numerical modelling.

In summary, we experimentally demonstrate \ZB{} in planar microcavity and microcavity honeycomb lattices. Unlike previous demonstrations the Fabry-Perot cavity design allows direct visualisation of the complex spinor field inside the device. Lattice structures allow building photonic analogs to important physical system, such the photonic graphene we study here. In exciton-polariton lattices, identical to our structure apart from smaller energy detuning between photons and quantum well exciton, interparticle interactions and high sensitivity to magnetic fields can easily be added. This work then opens the door to studying a very wide class of photonic analogs of relativistic systems with particle interactions, time-reversal symmetry breaking and dissipative effects.


\section*{Data availability}
The data supporting the findings of this study are available from the corresponding author upon reasonable request.

\section*{Acknowledgements}
The work of AO, AY and IAS was supported by Priority2030 Federal Academic Leadership Program. IAS acknowledges support from Icelandic Research Fund (Rannis), project No. 163082-051.

\section*{Conflict of interest statement}
All authors declare that there are no conflicts of interests.

\section*{Author Contributions}
DNK and IAS conceived the experiment. SL, CEW and PMW designed the experiment. SL performed the experiments with contributions from PUN. SL, PMW and AO analysed the experimental data. AO and AY developed the theoretical model and performed numerical simulations. PWM, SL, AY, AO and IAS wrote the manuscript and supplementary materials. All authors contributed to discussion of the data and discussion and revision of the manuscript.

\end{document}